# Towards decoding the coupled decision-making of metabolism and epithelial-mesenchymal transition in cancer

Dongya Jia[1,*], Jun Hyoung Park[2], Harsimran Kaur[3], Kwang Hwa Jung[2], Sukjin Yang[2], Shubham Tripathi[1,4,5], Madeline Galbraith[1,6], Youyuan Deng[1,7], Mohit Kumar Jolly[3], Benny Abraham Kaipparettu[2,8,*], José N. Onuchic[1,6,9,10,*], Herbert Levine[1,5,11,*]

[1]Center for Theoretical Biological Physics, Rice University, Houston, TX 77005, USA;

[2]Department of Molecular and Human Genetics, Baylor College of Medicine, Houston, TX 77030, USA;

[3]Centre for BioSystems Science and Engineering, Indian Institute of Science, Bangalore, Karnataka 560012, India;

[4]PhD Program in Systems, Synthetic, and Physical Biology, Rice University, Houston, TX 77005, USA;

[5]Department of Physics, Northeastern University, Boston, Massachusetts 02115, USA;

[6]Department of Physics and Astronomy, Rice University, Houston, TX 77005, USA;

[7]Applied Physics Graduate Program, Rice University, Houston, TX 77005, USA;

[8]Dan L. Duncan Comprehensive Cancer Center, Baylor College of Medicine, Houston, TX 77030, USA;

[9]Department of Chemistry, Rice University, Houston, TX 77005, USA;

[10]Department of Biosciences, Rice University, Houston, TX 77005, USA;

[11]Department of Bioengineering, Northeastern University, Boston, MA 02115, USA

**\*Correspondence:**

Dongya Jia (dyajia@gmail.com); Benny Abraham Kaipparettu (kaippare@bcm.edu); José N. Onuchic (jonuchic@rice.edu); Herbert Levine (h.levine@northeastern.edu)

**Abstract**

Cancer cells have the plasticity to adjust their metabolic phenotypes for survival and metastasis. During metastasis, a developmental program known as the epithelial-mesenchymal transition (EMT) plays a critical role. There is extensive cross-talk between metabolism and EMT, but how this leads to coordinated physiological changes is still uncertain. The elusive connection between metabolism and EMT compromises the efficacy of metabolic therapies targeting metastasis. In this review, we aim for clarifying causation between metabolism and EMT based on recent experimental studies and propose integrated theoretical-experimental efforts to better understand the coupled decision-making of metabolism and EMT.

**BACKGROUND**

Metastasis remains the leading cause of cancer-related deaths and thus efforts to interfere with metastasis and the related emergence of drug resistance remain of the highest priority. A developmental program referred to as the epithelial-mesenchymal transition (**EMT**) is often implicated in metastasis and acquisition of stemness which is typically associated with drug resistance [1]. During EMT, cobblestone-shaped epithelial cells lose their apical-basal polarity and cell-cell adhesion and become spindle-shaped with increased motility and invasiveness [1]. Crucially, cells undergoing EMT typically become more resilient in the face of stress, stress arising either from applied therapeutics and/or from being faced by a new microenvironment in a distant metastatic niche. Several recent review articles summarize the state-of-affairs concerning EMT phenomenology and EMT systems biology modeling [2,3].

Metabolic reprogramming (**MR**) is an emerging hallmark of cancer [4]. While normal cells mainly use oxidative phosphorylation (OXPHOS) for ATP production, cancer cells have been observed to rely primarily on glycolysis, irrespective of the presence of oxygen, referred to as the Warburg effect or aerobic glycolysis. It was hypothesized that the increased glycolytic activity in cancer was due to their defective mitochondria [5], but this has proven not to be the case. The proposed benefits of the Warburg effect include rapid ATP production, biomass synthesis and reactive oxygen species (ROS) balance [6]. Of note, several types of normal cells also exhibit glycolysis. For example, the adult stem cells that reside in hypoxic niches use glycolysis to maintain their self-renewal capacity [7]. Neural crest cells utilize aerobic glycolysis when undergoing EMT [8]. The observation supports the hypothesis that aerobic glycolysis is a physiological adaptive developmental program that is anomalously hyper-activated in cancer.

While the Warburg effect is a common phenomenon in cancer, the role of OXPHOS cannot be ignored. Increasing glycolytic activity in the primary tumor relative to normal cells does not necessarily mean its mitochondrial activity has to be suppressed. As one example, the pancreatic cancer PANC-1 cells exhibit significantly higher glucose oxidation activity relative to healthy pancreatic epithelial cells [9]. A meta-analysis of 31 cancer cell lines shows that the contribution of OXPHOS to ATP production ranges from 36% to 99% across cancer cell lines [10]. In short, mitochondrial respiration has demonstrated its critical role especially in metastasis [11–14].

Recent studies have witnessed significant advances in separately characterizing EMT and metabolic plasticity. It has been convincingly shown that neither EMT nor MR is a binary decision-making process, where cells can be only epithelial (E) or mesenchymal (M) or can utilize only glycolysis or OXPHOS. Instead, a more nuanced set of pictures has emerged. Cancer cells can exist along a spectrum of EMT states characterized by varying proportions of epithelial and mesenchymal traits [15–17]. Stable hybrid E/M phenotypes co-expressing E-cadherin and vimentin and exhibiting collective migratory behavior have been identified at the single-cell level and have been argued to be of critical importance to metastasis [16,18]. In addition, cancer cells can mix and match different aspects of energy production and resource utilization and acquire a hybrid metabolic phenotype characterized by high glycolysis and high OXPHOS [12,13,19,20]. This flexibility has proven problematic for the development of effective drug treatment, as tumor cells can readily adapt and continue to flourish even when putatively crucial pathways are blocked. Adding to the flexibility, cancer cells can exhibit metabolic coordination as shown during collective invasion wherein leader cells use more OXPHOS and the follower cells use more glycolysis [21].

It is reasonable to expect that changing motility phenotypes would necessitate altered cellular bioenergetics and thereby metabolism. Indeed, extensive regulatory cross-talk between EMT and MR has been demonstrated [22]. Yet, the cause-and-effect relationship remains elusive. Here, we focus on elucidating EMT-MR causation, building up a framework to think about the EMT-MR connection, and discuss how systems biology approaches can be developed to rationalize the EMT-MR connection.

**HOW DOES EMT AFFECT METABOLISM?**

EMT is a multi-dimensional transformation process involving changes in cellular mechanics and biochemical signaling, that are fine-tuned by the underlying epigenetic landscape [23]. These changes can be instigated by microenvironmental signals such as TGF-β, Notch ligands such as Delta and Jagged, inflammatory agents such as IL6 and TNF-α (acting through NF-κB), interactions with the extracellular matrix (ECM), and possibly by mutational events. These microenvironmental signals, initiating a type of stress response in the cancer cells, eventually impinge on a network involving EMT-inducing transcription factors (EMT-TFs) such as SNAIL and ZEB, and microRNAs (miRNAs) such as miR-200 and miR-34. Surprisingly, the details of these interactions seem to be context-dependent, indicating an elaboration of possible mechanisms en route to a relatively conserved set of phenotypic changes. Of note, the EMT-inducing signals and EMT-TFs also play critical roles in regulating cancer invasiveness in non-carcinoma contexts such as glioblastoma [24]. In the following, we discuss how these most important nodes of EMT regulation impinge upon metabolic reprogramming.

*EMT-inducing signals and metabolism*

Multiple microenvironmental signals that can trigger EMT can also reprogram metabolism. The EMT-inducing signal that best exemplifies such a dual effect as causing EMT and altering metabolism is TGF-β. Indeed, TGF-β has been widely used to induce EMT in a plethora of cancer types [23]. Concerning metabolism, TGF-β can promote glycolysis [24] and conversely, up-regulate fatty acid β-oxidation (FAO) as well to meet energy needs associated with EMT and motility [22,25]

TGF-β signaling and glycolysis: TGF-β signaling up-regulates glycolysis by increasing expression of both glucose transporters and glycolytic enzymes (Fig. 1). For example, TGF-β can increase the mRNA levels of glucose transporter 1 (GLUT1), hexokinase 2 (HK2) and lactate dehydrogenase A (LDHA) in glioblastoma [24]. The glycolysis-enhancing effect of TGF-β has also been reported in the PANC-1 cells that exhibit increased expression of PFKFB-3, an allosteric activator of the glycolytic enzyme PFK-1, followed by increased glucose uptake and lactate production upon TGF-β treatment [26]. Intriguingly, OXPHOS activity remains significant in these PANC-1 cells [27], indicating the acquisition of a hybrid metabolic phenotype. Perhaps, the increase of glycolysis during TGF-β induced EMT is connected to the well-known relationship between EMT and stemness [28,29] and the well-known tendency for stem cells to use this form of energy production [7]. This then leads to the question of whether these metabolic changes are in fact necessary for EMT. At least in some cases, the answer appears to be yes. As one example, PFKFB-3 knockdown suppresses SNAIL and reduces EMT-dependent invasiveness of PANC-1 cells [26]. A detailed discussion of the effects on EMT of induced metabolic changes is presented below.

TGF-β signaling and FAO: Aside from up-regulating glycolysis, TGF-β signaling can promote FAO. This will increase the rate of energy production as compared to the pre-EMT baseline. Non-small cell lung cancer (NSCLC) A549 cells, a cell line that uses basal glycolysis and OXPHOS [30], exhibits increased FAO along with EMT upon TGF-β treatment [22]. TGF-β treatment decreases the master lipogenic regulator ChREBP, and the reduction of ChREBP decreases expression of FASN and induces FAO. Again, the connection is two-sided; FASN knockdown is sufficient to induce EMT in vitro and can promote metastasis in vivo [22]. Moreover, TGF-β signaling can promote FAO by increasing fatty acid uptake via CD36 [25]. Interestingly, enhanced FAO enriches the cellular pool of acetyl coenzyme A (acetyl-CoA) which can increase Smad2 acetylation and strengthen the TGF-β signaling pathway. Therefore, there can be a mutual excitatory feedback loop between TGF-β signaling and FAO.

Put together, along with EMT induction, TGF-β can enhance both glycolysis and FAO, therefore it has the potential to induce cells to increase either glycolysis or FAO or both. Depending on how significant the increases of glycolysis and FAO are, the coupling between EMT and MR due to TGF-β stimulation can be context-dependent.

*EMT-TFs and metabolism*
In addition to investigating how external EMT-inducing signals couple to MR, one can study how specific EMT-TFs (SNAIL, SLUG, TWIST, ZEB etc.) cause specific metabolic changes whether they were stimulated by TGF-β or by other pathways (Fig. 1). This also allows for a better understanding of the EMT status of a given cell line and its expected metabolism.

SNAIL can down-regulate OXPHOS and up-regulate glycolysis. In a panel of basal-like breast cancer cell lines, SNAIL directly represses FBP1, a rate-limiting enzyme in gluconeogenesis [31]. FBP1 repression leads to reduced oxygen consumption and ROS production, and increased glycolysis and biomass synthesis. Indeed, the repression of FBP1 is required for SNAIL-induced EMT as ectopic expression of FBP1 abrogates the decrease of E-cadherin and EMT-related morphological change upon SNAIL induction. Conversely, knockdown of SNAIL can increase FBP1 expression.

In addition to SNAIL, SLUG and TWIST can also inhibit mitochondrial respiration and activate glycolysis. Overexpression of SLUG or TWIST in the luminal A breast cancer MCF7 and in the immortalized but non-tumorigenic MCF10A cells leads to decreased mitochondrial respiration and reduced mitochondrial mass; this is due to suppression of succinate dehydrogenase (SDH), an enzyme involved in the TCA cycle and the electron transport chain [32]. Notably, both the MCF7 and the MCF10A cells exhibit active basal OXPHOS [33]. An alternative mechanism underlying TWIST-mediated metabolic change in MCF10A cells relies on the PI3K/AKT/mTOR signaling that leads to an increase of pyruvate kinase M2 (PKM2) and LDHA [34].

ZEB1, a EMT-TF considered to be more critical for sustained EMT response relative to SNAIL/TWIST in certain scenarios [35,36], is crucial for cancer metabolic plasticity [35]. In the KPC mouse model of pancreatic cancer, unlike the parental KPC cells, ZEB1-knockout KPC cells fail to up-regulate their glycolytic activity to compensate for the decreased OXPHOS upon oligomycin treatment [35]. ZEB1 can promote glucose uptake by transcriptionally activating glucose transporter 3 (GLUT3) during EMT [37].

Thus, the EMT-TFs (SNAIL, SLUG, TWIST and ZEB1) exhibit consistent activation of glycolysis and inhibition of glucose-based OXPHOS acting via alteration of enzyme gene expression. Cells may or may not compensate for this rerouting of glucose by increasing FAO. However, there appear to be exceptions to this general behavior. Under oxidative stress, SNAIL can suppress glycolysis in MCF7 and the triple negative breast cancer (TNBC) MDA-MB-231 cells by transcriptionally inhibiting the glycolytic enzyme PFKP [38]. Down-regulation of PFKP diverts the glucose flux from glycolysis to the pentose phosphate pathway (PPP) to generate

NADPH for survival under oxidative stress. Another exception appears to arise during EMT of mouse breast cancer 4T1 cells; the circulating tumor cells (CTCs) presumably formed by 4T1 cells (4T1-CTCs) that have undergone EMT exhibit the highest OXPHOS and ATP production, mediated by PGC-1α; these cells retain a level of glycolytic activity similar to both the primary tumor and lung metastases formed by 4T1 [12]. It would be interesting to investigate whether the increased OXPHOS in 4T1-CTCs is due to increased FAO by evaluating the effect of FAO inhibitors such as ETX on the metabolic activity of 4T1-CTCs.

*EMT-associated-non-coding RNAs and metabolism*

The ncRNAs including both miRNAs, (e.g., miR-200 and miR-34) and long non-coding RNAs (lncRNAs, e.g., NEAT1 and ANRIL) comprise an important layer of post-transcriptional regulation of EMT. Both miRNA and lncRNA can function as either EMT-inducers (such as miR-10b [39] and HOTAIR [40]) or EMT-suppressors (such as miR-200 [41] and TUSC7 [42]).

EMT-associated miRNAs and metabolism: Generally speaking, the miRNAs that maintain an epithelial phenotype tend to repress glycolysis; conversely, the miRNAs repressing EMT can repress FAO as well. For example, the miR-200 family and the miR-34 family function as critical gatekeepers of the epithelial phenotype [41]. The miR-200 family directly targets ZEB, and the miR-34 family directly targets SNAIL. Both miR-200 and miR-34 can target LDHA to repress glycolysis [43,44]. miR-33, that inhibits EMT by targeting SNAIL and ZEB [45], can inhibit FAO by targeting genes encoding CPT1A and 5' AMP-activated protein kinase (AMPK) α [46]. The results here are reminiscent of the effect of TGF-β on both glycolysis and FAO when inducing EMT.

EMT-associated lncRNAs and metabolism: In general, the EMT-promoting lncRNAs can up-regulate glycolysis and FAO. For example, HOTAIR that can induce EMT by up-regulating EZH2 and sequentially EMT-TFs (SNAIL, ZEB and TWIST) [40], can enhance glycolysis by increasing the expression of GLUT1 by activating mTOR signaling [47]. Another example is the EMT-promoting lncRNA NEAT1 which can promote FAO through the activation of ATGL-PPARα signaling [48]. Again, we see the dual nature of the EMT effect on metabolism, able to increase both glycolysis and FAO with an adjustable balance to meet the cell's new requirements.

In summary, we have described how EMT-inducing signals (i.e. TGF-β), EMT-TFs and EMT-promoting lncRNAs can enhance both glycolysis and FAO, and suppress glucose-based OXPHOS. We have also described how EMT-suppressing miRNAs can repress both glycolysis and FAO, acting via altering the expression of metabolic transporters and enzymes. We have also mentioned several reciprocal cases where metabolic changes act as a causative agent for promoting or inhibiting EMT. We next look at this question more systematically.

**HOW DOES CANCER METABOLISM AFFECT EMT?**

Cancer cells are endowed with the metabolic plasticity to exploit the surrounding nutrients to adjust their metabolic activity including glycolysis, mitochondrial respiration, fatty acid metabolism, glutamine metabolism [19]. In this section, we will discuss how each of these processes can shape EMT, starting with the metabolic processes happening mainly in the cytosol, followed by those in the mitochondria (Fig. 1).

*Glycolysis can promote EMT*

Emerging evidence supports the notion that increased glycolysis can facilitate EMT. Cancer cells exhibit elevated glucose uptake relative to normal cells. This can be achieved via the overexpression of GLUT1 [49]. Overexpression of GLUT1 can induce matrix metalloproteinase 2 (MMP-2), a molecule typically up-regulated during EMT and can drive cell migration and invasion in multiple types of cancer [49]. Next, we will discuss how glycolytic enzymes facilitate EMT per their orders of appearance in the glycolytic pathway.

Up-regulation of HK2, the enzyme involved in the first rate-limiting step of glycolysis, can increase glycolysis and enhance metastasis of PANC-1 cells [50]. Consistently, down-regulation of HK2 results in decreased glycolysis and suppressed EMT [51]. The enzyme GPI, besides converting glucose-6-phosphate to fructose-6-phosphate, can act as a cytokine. When secreted by tumor cells, GPI is often referred to as an autocrine motility factor that can induce EMT via ZEB1/2 in a NF-κB-dependent manner [52]. PFKFB-3 or GAPDH, when inhibited, can inhibit

EMT via the downregulation of SNAIL [26,53]. PKM2 can induce EMT via its nuclear translocation that leads to transcriptional suppression of CDH1 [54]. Intriguingly, nuclear PKM2 can also activate β-catenin that enhances MYC activity, resulting in an increased expression of PKM2, LDHA and GLUT1, thus forming a self-reinforcing feedback loop to strengthen the connection between glycolysis and EMT [55].

At the last step of glycolysis, PDK1, the enzyme that inactivates pyruvate dehydrogenase (PDH) to avoid the conversion of pyruvate to acetyl-CoA to fuel the TCA cycle, promotes EMT by activating NF-κB signaling in gastric cancer [56]. Down-regulation of PDK1 reverts EMT and eliminates cisplatin resistance, a common EMT-associated trait, in ovarian cancer [57]. Interestingly, another isozyme of PDK - PDK4 - exhibits an opposite effect on EMT. PDK4 overexpression partially blocks TGF-β-induced EMT in lung cancer [58]. PDK4 inhibition is sufficient to induce EMT and enables erlotinib resistance. This result is reminiscent of the anti-metastasis effect in breast cancer obtained by suppressing pyruvate carboxylase (PC), which catalyzes the carboxylation of pyruvate to oxaloacetate to replenish the TCA cycle [59]. Altogether, these results suggest that lung and breast cancer cells that have undergone EMT may still need the diversion of some glucose to the TCA cycle, presumably because the TCA activity is needed to provide citrate to enrich the cellular pool of acetyl-CoA to facilitate EMT, as discussed in the next section. It would be interesting to investigate the dependence of this finding on glutamine transport, which can act as an alternate source of acetyl-CoA.

Finally, LDHA that converts pyruvate to lactate, the end product of glycolysis, can promote EMT by up-regulating ZEB2 [60]. The lactate produced and secreted by tumor cells, can lower the extracellular pH and convert the inactive extracellular TGF-β to its active form, thus promoting EMT [61]. Meanwhile, the secretion of lactate results in elevation of intracellular pH that activates Wnt signaling which can also activate EMT [62]. In summary, glucose transporters, essentially most glycolytic enzymes except for PDK4 and lactate accumulation, consistently promote EMT across cancer types.

*FAO can promote EMT*

In addition to glycolysis, fatty acid uptake and FAO can induce EMT/metastasis. TCGA pan-cancer analysis reveals that samples with a higher EMT score have a higher expression of CAV1 and CD36, both involved in fatty acid uptake [63]. A follow-up study showed that elevated fatty acids uptake via CD36 activates TGF-β and Wnt signaling pathways and EMT induction [64]. Enhanced FAO can promote metastasis presumably by activating oncogenic pathways such as Src [14] to promote EMT [65]. Consistently, blocking FAO by silencing CPT1A represses metastasis [14].

*Glutaminolysis and EMT*

Beyond glycolysis and FAO, glutaminolysis emerges as another important metabolic hallmark of EMT [66]. Among the enzymes involved in glutaminolysis, the role of glutaminase (GLS) in the regulation of EMT has been most widely reported. GLS includes two isoforms - the kidney-type GLS1 and the liver-type GLS2. Although both GLS1 and GLS2 can convert glutamine to glutamate, GLS1 functions as an EMT-inducer while GLS2 functions as an EMT-suppressor. GLS1, when knocked out, blocks TGF-β-induced EMT in MCF7 cells and represses metastasis in vivo [67]. GLS2 can bind to and stabilize Dicer to increase expression of miR-34 to repress EMT [68]. Consistently, the benign breast HMLE cells undergoing EMT exhibit increased GLS1 and decreased GLS2 [69]. In addition to GLS, GDH, which converts glutamate to α-KG, promotes EMT potentially via the STAT3 signaling [70]. Interestingly, reduced intracellular glutamine level, due to increased glutamine export by up-regulating the glutamine transporter SLC38A3, can suppress EMT by activating the PDK1/AKT signaling [71].

*Mitochondrial regulation of EMT epigenetics*

Mitochondrial enzymes and metabolites can promote EMT via their effects on epigenetic modifications. Cancer cells undergoing EMT exhibit epigenetic plasticity encompassing both DNA and histone modifications [23]. Modification enzymes are mainly classified into three categories - "**writers**" (e.g., histone acetyltransferases (HATs)) that attach molecular modification to chromatin or DNA , "**erasers**" (e.g., ten eleven translocation (TET)) that remove the modifications and "**readers**" (e.g., bromodomain containing protein 4 (BRD4)) that recognize acetylated or methylated residues [72]. Recently, an increasing number of reports have demonstrated that the metabolites generated in mitochondria can serve as crucial cofactors or substrates (referred to as "**ink**") - acetyl-CoA, α-KG, 2-HG, succinate, fumarate, *etc*. that directly impact the activity of histone modification enzymes [73].

Acetyl-CoA is a substrate for the "writer" HATs to perform acetylation which generates an euchromatin state of the target genes that encode EMT-TFs to induce EMT. Acetyl-CoA accumulation can increase H3 acetylation in the promoter region of TWIST2 and consequently induces TWIST2 expression and activates EMT, as shown in HCC [74]. It is important to note that in addition to regulating histone acetylation and resultant epigenetic effects, acetyl-CoA can also function as a key cofactor for post-translational modification of key proteins involved in EMT. For example, with a high level of histone acetylation, the "reader" BRD4 can bind to and stabilize acetylated Snail and consequently promotes EMT, as shown in gastric cancer [75]. Elevated acetyl-CoA level can also increase Smad2 acetylation to facilitate EMT, as shown in breast cancer [25].

α-KG, which is generated from citrate during the TCA cycle or by glutaminolysis, is a substrate for the 'eraser' TETs [76]. The accumulation of α-KG can increase TET activity to cause DNA demethylation of miR-200 and as a result, miR-200 is activated and EMT is inhibited. [77] In addition, the accumulation of α-KG can stimulate nitric oxide (NO) production, and up-regulate NO-sensitive miR-200.

Due to the structural similarity with α-KG, 2-HG, succinate and fumarate perform as competitive inhibitors of α-KG-dependent demethylases [78–80]. In various human cancer types, gain-of-function mutation of both IDH1 (cytosol) and IDH2 (mitochondria) are frequently observed. As a result, 2-HG accumulates and can promote EMT through increasing the H3K4 methylation in the promoter region of ZEB1 [81]. Increased levels of succinate and fumarate in cancer are often due to the loss-of-function mutation of SDH and FH, respectively. Succinate accumulation can cause DNA hypermethylation by inhibiting TET2, and subsequently promotes EMT [82]. Fumarate accumulation inhibits TETs-mediated DNA demethylation of miR-200ba429 and consequently, the repression of ZEB is relieved and EMT is activated [83].

Overall glucose, fatty acid, glutamine metabolism can both cause or halt EMT via alteration of the expression of genes or activity of proteins involved in EMT. While certain trends are more common, we need to recognize that there is no simplistic rule for how EMT needs to be coordinated with metabolism in a specific cell line or specific patient sample. Interestingly, this finding is analogous to the connection between EMT and stemness, where it has also been realized that cells at various stages of EMT can all become stem-like [29]; there is a most common behavior (namely that hybrid E/M cells are more conducive to de-differentiation) but the detailed behavior can be cell line and patient sample dependent.

**THE RISE OF THE HYBRID PHENOTYPE**

As has already been mentioned, cancer cells can acquire hybrid E/M and hybrid metabolic phenotypes. The hybrid E/M phenotype has been proposed to be the primary instigator of metastasis and acquisition of stemness [16,84,85]. The hybrid metabolic phenotype characterizes the highly metastatic 4T1 cells [13] and breast cancer stem cells (BCSCs) [86]. As both hybrid E/M and hybrid metabolic phenotypes have been separately proposed to primarily account for high metastatic potential and cancer stemness, there ensues an interesting question, whether the hybrid metabolic phenotype typically characterizes the hybrid E/M phenotype. The answer seems to be, in many cases, yes.

Let us first investigate CTCs. Breast cancer-derived CTCs exist in two forms in vivo - small clusters of CTCs (>90%) and single CTCs (<10%) [87]. The clusters of CTCs arise from cases where tumor cells migrate collectively and have been attributed to the hybrid E/M phenotype [84]. The 4T1-CTCs (CTCs formed by implanted 4T1 cells) exhibit highest OXPHOS, and similar glycolysis relative to both the primary tumors and lung metastases formed by 4T1, indicating the acquisition of a hybrid metabolic phenotype [88]. Putting it all together, it is likely that the 4T1-CTCs contain largely clusters of CTCs that acquire both a hybrid E/M phenotype and a hybrid metabolic phenotype. Supporting the increased OXPHOS in CTCs, the antioxidation regulator NRF2 reaches its maximum in the hybrid E/M phenotype relative to E and M [89].

Similar ideas are emerging from the consideration of CSCs. BCSCs contain two subpopulations - the CD44$^{high}$/CD24$^{low}$ more-mesenchymal-like BCSCs (M-BCSCs) and the ALDH$^+$ hybrid E/M-BCSCs (sometimes referred to as the more-epithelial-like BCSCs) [86,90]. Relative to M-BCSCs, the hybrid E/M-BCSCs have higher OXPHOS and similar glycolysis, i.e., have acquired a hybrid metabolic phenotype. The simplest way of thinking about this data is that differentiated epithelial cells mostly use OXPHOS, as they transition to hybrid E/M-like stem cells they turn on significant glycolysis but retain their OXPHOS emphasis, and as they further transition to M-like stem cells they decrease their OXPHOS levels. This quiescent phenotype is then the most stem-like state, consistent with the notion that glycolysis is useful for enabling cellular plasticity [7]. The cells eventually transition to differentiated mesenchymal cells, leaving stemness behind and presumably increasing OXPHOS activity and again relying less on glycolysis (Figure 2).

We have discussed scenarios wherein epithelial cancer cells can either up-regulate glycolysis while maintaining OXPHOS [27], or up-regulate OXPHOS while maintaining glycolysis [12], after EMT is initiated. Relative to normal cells, the elevated overall metabolic activity in cancer is made possible by the increased capacity of cancer cells to import more glucose [49], fatty acid [63], and glutamine [66]. These increases can be potentially further strengthened during EMT by TGF-β [24] and by EMT-TFs [37]. The increased metabolic activity of cancer cells undergoing EMT is further backed up by the observation that the cytosol ATP levels reach maximum in the middle of EMT [91]

Therefore, a detailed consideration of both the epithelial and mesenchymal axis and the axis measuring glycolysis and OXPHOS leads to the hypothesis that in many cancer cells a hybrid metabolism can characterize a hybrid E/M phenotype (hybrid-hybrid). We should be cautiously optimistic about the hybrid-hybrid connection because the measured hybrid metabolic phenotype of the 4T1-CTCs and BCSCs can be consistent with single-cell hybrids but also with a mixture of varying phenotypes. As methods are being developed to study metabolomics at the single-cell level to complement existing single-cell RNA-seq data, this relationship will come into clearer focus as will the sequence of temporal events leading to these correlated changes in different aspects of cellular physiology.

**TO SOLVE THE EMT-MR PUZZLE: SYSTEMS BIOLOGY**

To further elucidate EMT-MR causation, we need to systematically investigate the dynamics of the coupled decision-making networks of EMT and metabolism. To that end, systems biology approaches, which view living organisms as being composed of interacting dynamical networks containing nodes (genes, proteins and metabolites etc.) and links (gene regulation, protein-protein interaction and biochemical reactions etc.) should be developed and applied (Fig. 3).

There are two main categories of systems biology models, bottom-up and top-down. The bottom-up approach often starts with a relatively well-defined molecular network and formulates mathematical models to simulate its dynamics; the goal is to better explain observed experimental results and to generate predictions to guide new experiments. The top-down approach often starts from omics data from which phenomenological models are inferred to capture the data. Both bottom-up and top-down approaches have been applied to EMT and metabolism, albeit independently. As we will argue, these separate studies need to be combined to gain an integrated view of EMT-MR coupling.

To simulate EMT, mechanism-based mathematical models have been developed to analyze the dynamics of a proposed core regulatory network consisting of two mutually inhibitory feedback loops - SNAIL/miR-34 and ZEB/miR-200 [36]. This study predicts the acquisition of a stable hybrid E/M state as a possible end point of EMT. To simulate EMT dynamics at a relatively large scale, the Boolean network modeling approach and algorithms such as random circuit perturbation (RACIPE) have been applied, by which multiple hybrid E/M states have been characterized [92,93]. To identify the sequential gene states during EMT, statistical approaches such as Bayesian data assimilation, have been employed [94]. A detailed review of various modeling approaches in elucidating EMT dynamics can be found in [2,3].

To study metabolism, constraint-based methods including the flux-balance analysis (FBA) based on mass conservation, have been the most widely used approach [95]. FBA is well suited to quantify the steady-state metabolic flux distribution and can be adapted to characterize effects of perturbations in a computationally inexpensive way. As one example, FBA has been applied to identify synthetic lethal metabolic gene pairs, proposing targeting which can specifically repress cancer cells while leaving normal cells intact [96]. However, in its base form, FBA does not incorporate genetic regulation, and cannot determine metabolite concentrations [95]. To investigate the crosstalk between genetic regulation and metabolic pathways, a mechanism-based mathematical model coupling the metabolic fluxes (glucose oxidation, glycolysis and FAO) with the AMPK/HIF-1 genetic circuit has been proposed [20]. The study reveals a direct association between AMPK/HIF-1 and OXPHOS/glycolysis, illustrates how cancer cells switch between different metabolic phenotypes, predicts the existence of a stable hybrid metabolic phenotype and a metabolic inactivity phenotype which have been confirmed in TNBC and melanoma cells respectively [20,97]. These studies serve as examples of how systems biology approaches can be formulated and applied to probe cancer metabolism. A detailed review of the mathematical models of metabolism can be found in [98].

Future progress on using computational models to elucidate the coupled decision-making of EMT-MR must combine these independent studies after a careful assessment of their cross-talk. Some early efforts have been made toward this end. To characterize the metabolic change in isogenic epithelial and mesenchymal cells, constraint-based models have been formulated to analyze the metabolic flux in the normal-derived basal-like breast epithelial cell line D492 and its mesenchymal counterpart D492M. This study predicts that D492 acquires both higher glycolysis and higher mitochondrial ATP production relative to D492M, which was then validated by measuring OCR and ECAR [99]. Both D492 and D492M cell lines exhibit some stem-like properties and the difference in their metabolic features is reminiscent of those between the hybrid E/M-BCSCs and M-BCSCs [86]. It would be interesting to apply EMT scoring metrics [100] to quantify the EMT status of D492 to determine the extent to which it contains cells with a hybrid E/M phenotype.

One very recent attempt to achieve a better understanding of the correlated temporal changes in EMT status and metabolism relies on computing a probabilistic landscape for a regulatory network integrating the decision-making modules of metabolism, EMT and metastasis. Given a regulatory network, the landscape approach can characterize the basins of attractions corresponding to stable cell phenotypes, and approximate transition paths to predict the temporal orders of events. With that in mind, Kang et al. proposed a path to metastasis that epithelial cells first reprogram their metabolism by up-regulating HIF-1, then transition to mesenchymal by up-regulating ZEB followed by transitioning into a metastatic state with the up-regulation of BACH1 [101]. The result indicates that, as we have argued, metabolic reprogramming is necessary for EMT and interfering with metabolic changes can intercept EMT at a very early stage. Hopefully, this early attempt will spur the systems biology community to stop thinking of EMT and metabolic plasticity as uncoupled modules; as we have seen they are strongly intertwined.

## CONCLUDING REMARKS

As both EMT and metabolism are multi-dimensional processes involving stable hybrid phenotypes, it can be misleading when considering the EMT-MR connection if we simply classify a cancer cell line as purely epithelial or mesenchymal or imagine that genetic or pharmacological perturbations have an all or nothing EMT-related consequence. Similarly, we cannot label a cell's energy metabolism as being glycolytic or OXPHOS without having a detailed quantitative evaluation of both processes. Therefore, quantifying both EMT and metabolism is critical for probing the EMT-MR connection. It is probable that as EMT proceeds to different stages, different metabolic processes become prominent in turn. It is possible that metabolic changes that accompany EMT precede some of the more obvious indicators of become more mesenchymal and indeed actively contribute to the activation of these indicators. By coupling the decision-making networks of EMT and metabolism, systems biology approaches help identify the association of specific EMT and metabolic states and the conditions enabling that association. The predicted EMT-MR connections can be ideally tested by simultaneously performing transcriptomics and metabolomics analysis of the same single cells and by experiments performing perturbations on EMT to identify metabolic change and vice versa. A synergistic use of theoretical and experimental efforts should be made to delineate the dynamical EMT-MR connection map.

**Additional Information**

**Acknowledgements:** Not applicable.

**Authors' contributions:** D.J., B.A.K., J.N.O. and H.L. conceptualized the review. D.J., J.H.P., H.K., K.W.J., S.Y., S.T., M.G., Y.D. and M.K.J. contributed to writing early drafts of the paper. D.J., B.A.K., J.N.O. and H.L. revised the final version of the paper.

**Ethics approval and consent to participate:** Not applicable.

**Consent for publication:** Not applicable.

**Data availability:** Not applicable.

**Competing interests:** The authors declare no conflict of interest.

**Funding information:** This work is supported by the National Science Foundation (NSF) Center for Theoretical Biological Physics (NSF PHY-2019745) and NSF grants nos. PHY-1605817, PHY-1522550, PHY- 1842494, and CHEM-1614101. J.N.O. is a CPRIT Scholar in Cancer Research. D.J. is supported by a training fellowship from the Gulf Coast Consortia, on the Computational Cancer Biology Training Program (CPRIT grant no. RP170593). B.A.K. is supported by NIH grants nos. CA253445, CA234479, DK117001 and CA235113 and DOD grant no. W81XWH-18-1-0714. M.K.J. is supported by Ramanujan Fellowship awarded by Science and Engineering Research Board (SERB), Department of Science and Technology, Government of India (SB/S2/RJN-049/2018).

**FIGURE LEGENDS:**

**Figure 1 Cross-talk between EMT and metabolism in cancer.** The purple solid arrows represent metabolic fluxes. The black arrows/bar-headed arrows represent transcriptional regulation. The black dotted bar-headed arrows represent ncRNA-mediated regulation. The brown arrows/bar-headed arrows represent epigenetic regulations. The purple dotted arrows represent the transportation of molecules that mediate the EMT-MR interaction. EMT-inducing signals, EMT-TFs and EMT-associated ncRNAs can directly regulate the transcription or translation of metabolic enzymes and transporters. In turn, the metabolic intermediates can facilitate epigenetic modification of EMT-associated genes or proteins. GLUT: glucose transporter; FATP: fatty acid transporter protein; MCT: monocarboxylate transporter; LDHA: lactate dehydrogenase A; PDH: pyruvate dehydrogenase; PC: pyruvate carboxylase; OAA: oxaloacetic acid; α-KG: alpha-ketoglutarate; CPT: carnitine palmitoyltransferase; GDH: glutamate dehydrogenase; GLS: glutaminase; ncRNA: non-coding RNA; EMT-TF: EMT-inducing transcription factor.

**Figure 2 One compelling hypothesis regarding EMT-metabolism coupling during the acquisition of stemness.** Differentiated epithelial cancer cells depend on OXPHOS [9,10]. Upon undergoing EMT, the epithelial cells can increase their glycolytic activity as needed to acquire stemness and become hybrid E/M-like CSCs (OXPHOS[high]/glycolysis[high], proliferative). The hybrid E/M-like CSCs can either decrease their OXPHOS activity to transition into the mesenchymal-like CSCs (glycolysis[high], quiescent) [86,90] or decrease their glycolysis activity and lose stemness to transition into the differentiated mesenchymal cancer cells (OXPHOS[high]). Note that being labeled OXPHOS[high] does not determine the extent to which FAO is active, which could differ between different types of differentiated cancer cells, Quiescent mesenchymal-like CSCs (glycolysis[high]) can either increase their OXPHOS activity to become the hybrid E/M-like CSCs [86]or switch from glycolysis to OXPHOS and differentiate. The differentiated mesenchymal cancer cells may undergo dedifferentiation to become mesenchymal-like CSCs. It will be interesting to investigate whether differentiated epithelial cancer cells can directly transition into the mesenchymal-like CSCs without passing through hybrid E/M phenotypes. CSCs, cancer stem cells.

**Figure 3 A systematic pipeline to elucidate the EMT-MR connection.** Mathematical modeling approaches have provided critical insights into the dynamics of EMT and metabolism and predicted the acquisition of stable hybrid phenotypes by cancer. The AMPK/HIF-1/ROS model (bottom left) is an initial effort to understand how genetic regulation interacts with metabolic fluxes [86,102]. The miR-34/SNAIL/miR-200/ZEB model (bottom right) was used in the initial studies that proposed that hybrid E/M phenotypes can be an end point of EMT [36]. The predicted hybrid metabolic phenotype and hybrid E/M phenotype has been validated by several in vitro and in vivo studies [15,18,20]. With the rise of coupled hybrid metabolism with hybrid E/M phenotype, mathematical modeling studies that couple the decision-making networks governing metabolism and EMT can in principle provide a systematic view of the EMT-metabolism connection. With further advances in single-cell multiomics, acquisition of the transcriptomics and metabolomics profile of the same single cell can facilitate the understanding of EMT-metabolism coupling. Through an integrated mathematical modeling, data analysis and experimental studies, the nuanced EMT-metabolism connection can be gradually elucidated.

Figure 1

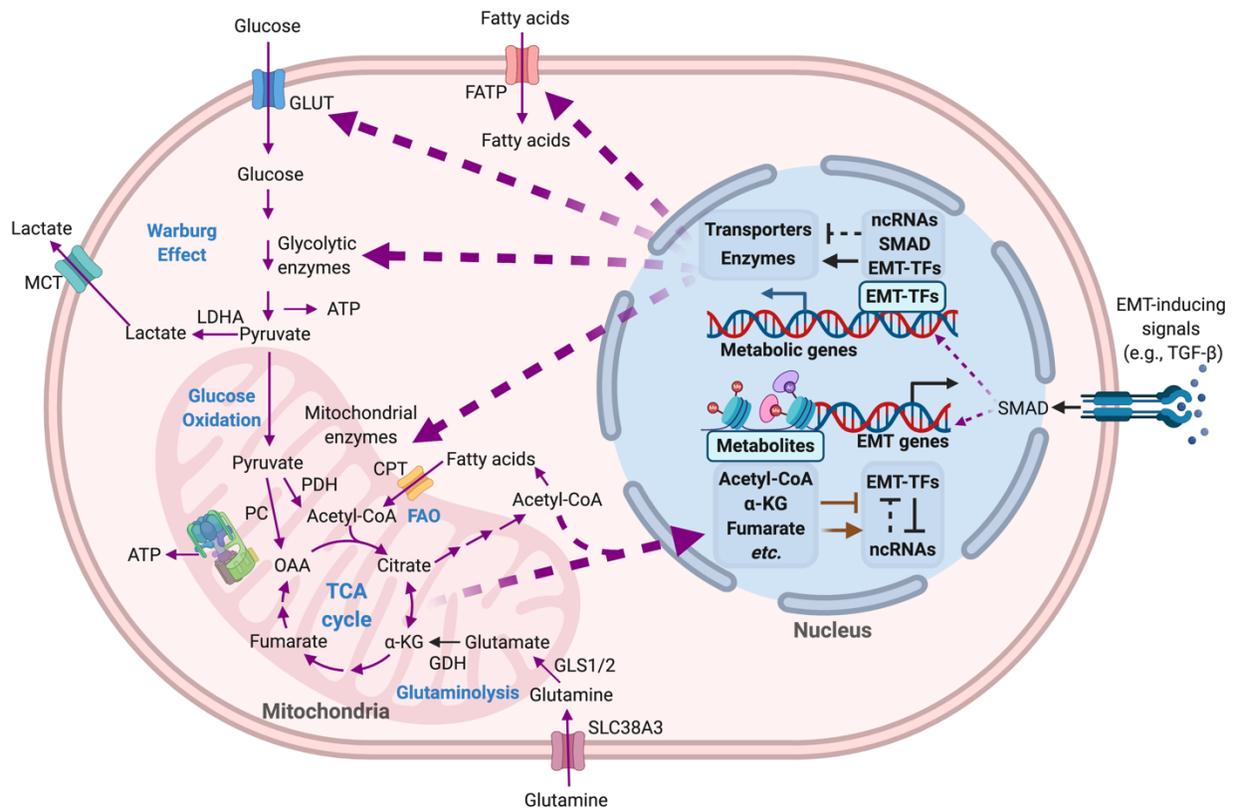

Figure 2

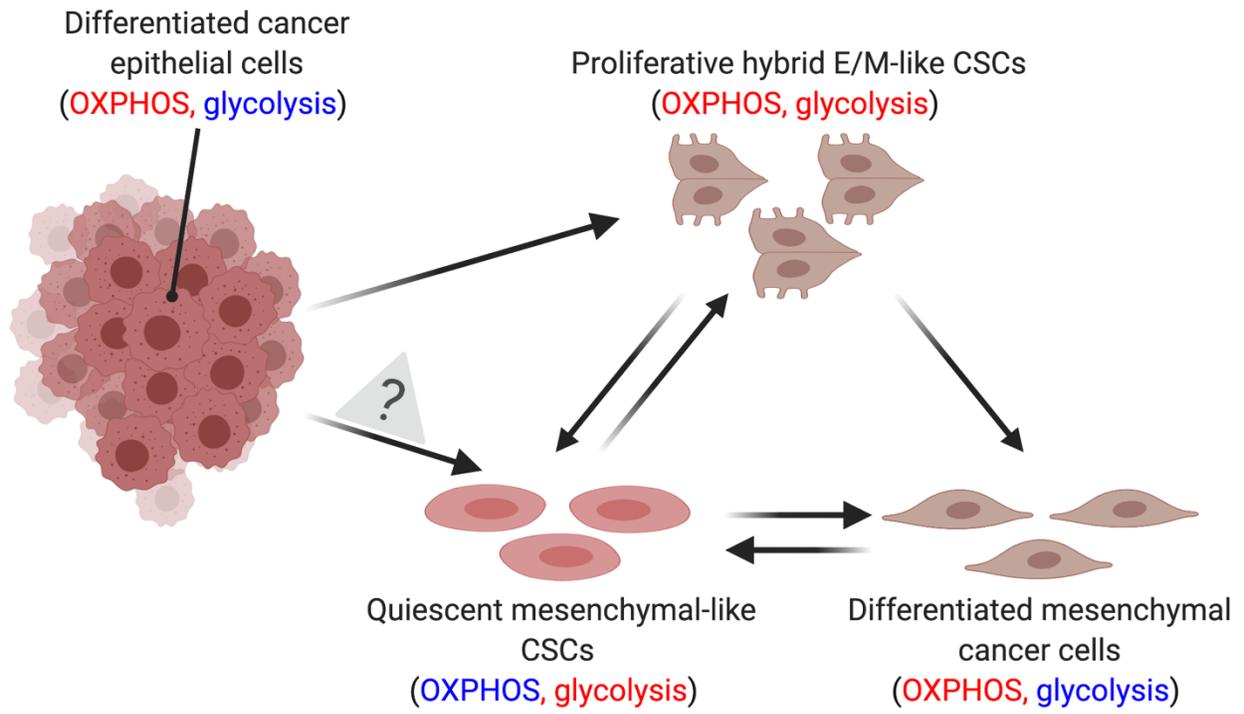

Differentiated cancer epithelial cells (OXPHOS, glycolysis)

Proliferative hybrid E/M-like CSCs (OXPHOS, glycolysis)

Quiescent mesenchymal-like CSCs (OXPHOS, glycolysis)

Differentiated mesenchymal cancer cells (OXPHOS, glycolysis)

Created with BioRender.com

Figure 3

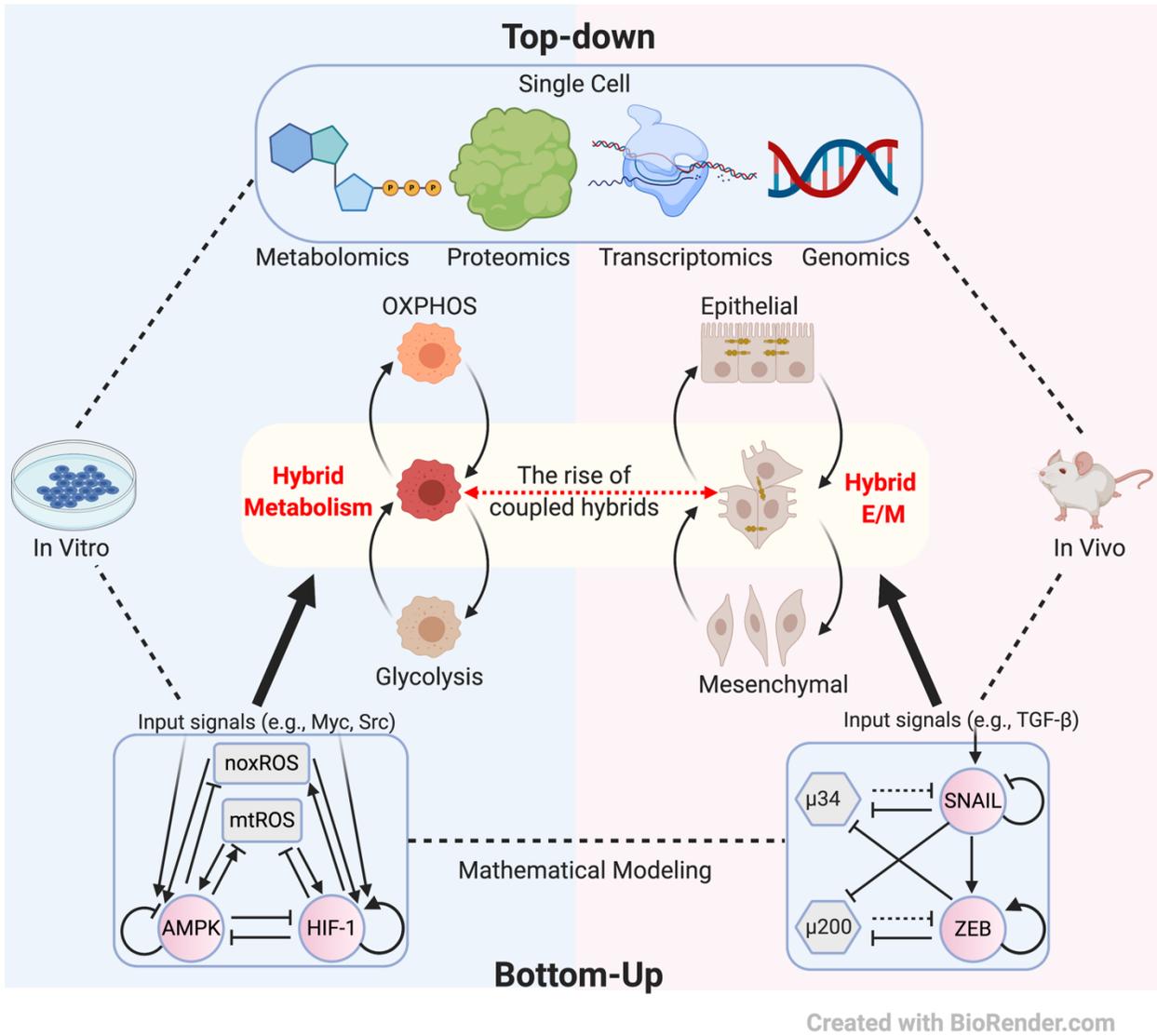